\documentclass[aps,pra,twocolumn]{revtex4}
\newcommand {\be}{\begin{equation}}
\newcommand {\ee}{\end{equation}}
\newcommand {\bey}{\begin{eqnarray}}
\newcommand {\eey}{\end{eqnarray}}

\usepackage{epsfig}
\usepackage{amsfonts}
\usepackage{amsmath}
\usepackage{mathbbol}

\usepackage{hyperref}

\begin{document}

\title{Information-based measure of nonlocality}
\author{Alberto Montina and Stefan Wolf}
\affiliation{Facolt\`a di Informatica, Universit\`a della Svizzera Italiana, 
Via G. Buffi 13, 6900 Lugano, Switzerland}

\date{\today}

\begin{abstract}
Quantum nonlocality concerns correlations among spatially separated systems that cannot 
be explained classically without communication among the parties. Thus, 
a natural measure of nonlocal correlations is provided by the minimal amount of 
communication required for classically simulating them. In this paper, we present a 
method to compute the minimal communication cost of parallel simulations, which we call 
nonlocal capacity,
for any general nonsignaling correlations. This measure turns out to have an important
role in communication complexity and can be used to discriminate between local and 
nonlocal correlations, as an alternative to the violation of Bell's inequalities. 
\end{abstract}

\maketitle

\section{Introduction}

The outcomes of measurements performed on spatially separate entangled systems can display 
nonlocal correlations that cannot be explained classically without some 
communication~\cite{bell}. In particular, one of the parties needs some information
on the measurement choice of the other party. These \emph{nonlocal correlations} can be 
used as an information-theoretic 
resource. For example, they can exponentially reduce the amount of communication
required to solve some distributed computational problems~\cite{cleve,buhrman}. 
Furthermore, for some tasks, the use of nonlocal correlations can make communication 
unnecessary, such as in pseudo-telepathy games~\cite{brassard}.
Some stronger-than-quantum nonsignaling correlations can even collapse the communication 
complexity in any two-party scenario. Indeed, the access to an unlimited number of 
Popescu-Rohrlich (PR) nonlocal boxes allows two parties to solve any communication complexity 
problem with the aid of a constant amount of classical communication~\cite{vandam}.
Nonlocal correlations have also a fundamental role in device-independent applications, such
as key agreement in cryptography~\cite{barrett0,acin0,scarani,acin2,acin3,masanes,hanggi}
and randomness amplification~\cite{colbeck,gallego}.

As the violation of a given Bell inequality is the signature of nonlocal correlations, 
a possible measure of nonlocality is the strength of this violation. However, since this 
quantity has no obvious relation with information, it does not necessarily 
provide a reliable measure as an information-theoretic resource. A more natural measure has been
employed in Refs.~\cite{maudlin,brassard,steiner,gisin1,gisin2} and relies on the very 
definition of nonlocality; nonlocal correlations require some communication to be classically 
simulated, thus the minimal amount of required classical communication can be used as 
a measure of the strength of nonlocality. This measure, which we call \emph{communication
complexity} of the nonlocal resource, provides an ultimate limit to the power 
of nonlocal correlations in terms of classical communication in a two-party scenario. 
Indeed, nonlocal resources cannot replace an amount of classical communication bigger 
than the associated communication complexity. As shown in Ref.~\cite{pironio},
the strength of the Bell inequality 
violation and the communication complexity of nonlocal resources turn out to be identical
if the average amount of communication is employed as measure of the
communication cost and the optimal inequality is taken for the given nonlocal 
correlations. In this paper, we mainly focus on the minimal 
asymptotic communication cost of parallel simulations in the asymptotic limit of infinite 
instances.
This quantity, which we call {\it nonlocal capacity}, turns out to be much
easier to be computed than its single-shot counterpart. Furthermore, 
tight lower and upper bounds on the minimal average communication cost 
are given in terms of the nonlocal capacity, as discussed later. Thus, the 
nonlocal capacity also gives tight bounds on the maximal violation
of the Bell inequalities. Alternative measures of 
nonlocality could use different resources as unit of nonlocality, such as nonlocal 
boxes~\cite{barrett,brunner}. For example, the strength of nonlocality could be
defined as the number of PR-boxes necessary to simulate the correlations.
However, no finite set of PR-boxes can simulate all bipartite
nonlocal correlations~\cite{dupuis,wolf}.

By definition, the computation of the nonlocal capacity is an optimization problem,
but it is not convex in its original form. This makes it very hard to find the global 
minimum, expecially when the set of allowed measurements is large.
In this paper, we show that the problem can be reduced to a convex minimization
problem, which can be numerically solved with very efficient algorithms~\cite{boyd}. Then, 
we discuss the relation with a previous work on the communication complexity of
channels in general probabilistic theories~\cite{montina1}. Finally, we illustrate
the method with a numerical example.

\section{Communication cost of nonlocal correlations}

\begin{figure}
\epsfig{figure=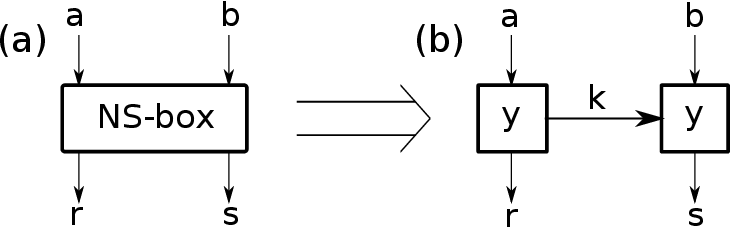,width=7.cm}
\caption{(a) Nonsignaling box with inputs $a$ and $b$ and outcomes $r$ and $s$. (b) Simulation
of the nonsignaling box through shared stochastic variable $y$ and communication of the 
variable $k$.}
\label{fig1}
\end{figure}
In this paper, we will discuss the general case of nonsignaling correlations, which
satisfy the minimal requirements of relativity and causality. Namely, the object 
that we will consider is a nonsignaling box, which is an abstract generalization
of the following quantum scenario. Two parties, say Alice and Bob, simultaneously 
perform a measurement on two spatially separate parts of an entangled system. In general,
Alice and Bob are allowed to choose among their respective sets of possible measurements.
We assume that Bob's set of measurements is finite, but arbitrarily large. For the sake
of simplicity, we also assume that Alice's set is discrete, although this is not
strictly necessary.
Let us denote by the indices $a$ and $b$ the measurements performed by Alice 
and Bob, respectively. The index $b$ takes a value in $\{1,\dots,M\}$, where $M$
is the number of measurements that Bob can perform. After the measurements, Alice 
gets an outcome $r$ and Bob an outcome $s$. The overall scenario is described by the 
joint conditional probability $P(r,s|a,b)$. This distribution satisfies the 
nonsignaling conditions
\be
\begin{array}{c}
\label{ns-conds}
\sum_{s} P(r,s|a,b)=\sum_{s} P(r,s|a,\bar b)\equiv P(r|a)\; \forall a,b,\bar b,r, 
\vspace{1mm} \\
\sum_{r} P(r,s|a,b)=\sum_{r} P(r,s|\bar a,b)\equiv P(s|b)\; \forall a,\bar a,b,s.
\end{array}
\ee
These conditions are implied by causality and relativity.
In the following discussion, we consider a more general scenario including
non-quantum correlations and we just assume that the joint conditional probability 
satisfies the nonsignaling condition. The abstract machine producing the correlated 
variables $r$ and $s$ from the inputs $a$ and $b$ will be called {\it nonsignaling 
box} (briefly, NS-box). The NS-box, schematically represented in Fig.~\ref{fig1}a, 
is identified with the conditional probability $P(r,s|a,b)$.

In general, a classical simulation of the joint distribution $P(r,s|a,b)$
requires some communication between the parties. We assume that only a one-way 
communication from Alice to Bob is allowed. The classical protocol is as follows 
(as illustrated in Fig.~\ref{fig1}b). Alice generates an outcome $r$ and a variable 
$k$ with probability $P(k,r|y\,a)$ 
depending on the variable $a$ and some stochastic variable $y$ shared with Bob
and generated with probability $\rho(y)$. The variable $k$ is sent to Bob.
Finally, Bob generates
an outcome $s$ with probability $P(s|y\,b\,k)$ depending on $y$, $b$ and
$k$. The protocol simulates the NS-box $P(r,s|a,b)$ if
\be
\label{cond_sim}
\sum_k\int dy P(s|y\,b\,k)P(k,r|y\,a)\rho(y)= P(r,s|a,b).
\ee
We define the {\it communication complexity} (denoted by ${\cal C}_{nl}$) of the NS-box as 
the minimal amount of communication $\cal C$ required for an exact simulation of 
the NS-box. 

There are different measures of \emph{amount of communication}. Here we employ the
entropic definition, although the presented results apply also to the case of 
average communication. Let us introduce the conditional probability
$$
P(k|y)\equiv \sum_{r,a} P(k,r|y,a)P(a)
$$
and the corresponding conditional Shannon entropy of the variable $K$ given $Y$ 
$$
H_{P(a)}(K|Y)\equiv-\int dy \rho(y)\sum_k P(k|y)\log_2 P(k|y),
$$
which depends on $P(a)$. We define the communication cost $\cal C$
of the simulation as the maximum, over the space of distributions
$P(a)$, of $H_{P(a)}(K|Y)$, that is,
\be\label{comm_cost}
{\cal C}\equiv\max_{P(a)} H_{P(a)}(K|Y)
\ee
(see also Refs.~\cite{montina1,montina2} and later discussion for the operational
interpretation). Note the abuse of notation in Eq.~(\ref{comm_cost}).
The maximization is performed with the respect to $P(a)$ as a function of $a$.
The argument of the function $P(a)$ is used to distinguish it from the other 
distributions, such as $P(s|y\,b\,k)$ and $P(k,r|y\,a)$. The same representation 
is used for the label of $H_{P(a)}(K|Y)$. For the sake of simplicity, we will use
this notation whenever the meaning is clear from the context.

The operational interpretation of $\cal C$ is provided by Shannon's source coding 
theorem and the {\it wrong code} theorem (Theorem~5.4.3 in Ref.~\cite{cover}). 
Given a compression code for $k$, let us denote by $L(a)$ the expected length of
the codeword of $k$ for a given input $a$ and by $\bar L[P(a)]\equiv \sum_a P(a)L(a)$ 
the expected length averaged over $a$ with the distribution $P(a)$. 
The interpretation of $\cal C$ is given by the following properties.
There is an optimal coding such
that the minimal worst-case \emph{expected} length $\max_a L(a)$ is equal to $\cal C$ 
up to one additional bit, that is,
\begin{equation}
\label{bound_La}
{\cal C}\le\max_a L(a)\le {\cal C}+1.
\end{equation}
In other words, for the optimal code, Alice needs to send not more than ${\cal C}+1$ 
on average for every choice of the input $a$ and this bound is strict for some input 
$a$ up to one bit. Furthermore, the optimal code minimizing $\max_a L(a)$ also
minimizes $\bar L[P(a)]$ for the worst-case distribution $P(a)$ and the minimum is
equal to $\cal C$ up to one bit.
It is worth to stress that the upper bound collapses to $\cal C$ 
if block-coding of $k$ is employed, as discussed later.
Let us prove Ineqs.~(\ref{bound_La}). Suppose that Alice and Bob employ
the optimal code minimizing  $\max_a L(a)$ and
Alice chooses the input $a$ according to the distribution $P(a)$ maximizing 
$H_{P(a)}(K|Y)$, denoted by $P_M(a)$. 
From Shannon's source coding theorem, we have that the expected length of
the codeword of $k$, $\sum_a P_M(a) L(a)$, is not smaller than 
${\cal C}=H_{P_M(a)}(K|Y)$. Thus,
\begin{equation}
{\cal C}\le \max_a L(a).
\end{equation}
Let us define the distribution
\begin{equation}
P_M(k|y)\equiv \sum_{r,a} P(k,r|y,a)P_M(a).
\end{equation}
From Shannon's theorem, it is clear that the optimal code minimizing $\bar L[P(a)]$
for the worst-case distribution $P(a)$ is the code that minimizes $\bar L[P_M(a)]$
and the minimum is equal to $\cal C$ up to one bit.
Now, we show that the optimal code minimizing $\bar L[P_M(a)]$
also minimizes $\max_a L(a)$ up to one additional bit. Namely, employing the 
optimal code for $P_M(a)$, the {\it wrong code} theorem implies that $\max_a L(a)$ is 
not bigger than ${\cal C}+1$. Indeed, if Alice generates $a$ according to a different 
distribution $P(a)$ and uses the code that is optimal for $P_M(a)$, the expected codeword 
length $\bar L[P(a)]$ is equal to the Shannon entropy $H_{P(a)}(K|Y)$
plus the relative entropy $D[P(k|y)||P_M(k|y)]$ up to an additional bit~\cite{cover},
where the relative entropy is~\cite{cover}
\begin{equation}
D[P(k|y)||P_M(k|y)]\equiv -\int dy \rho(y)\sum_k P(k|y)\log\frac{P(k|y)}{P_M(k|y)}.
\end{equation}
Thus, defining the quantity
\begin{equation}
\begin{array}{c}
{\cal C}[P(a)]\equiv H_{P(a)}(K|Y)+D[P(k|y)||P_M(k|y)]=  
\vspace{1mm} \\
-\int dy \rho(y)P(k|y)\log P_M(k|y),
\end{array}
\end{equation}
we have that
\begin{equation}
\label{Lbar}
\bar L[P(a)]\le {\cal C}[P(a)]+1.
\end{equation}
Let us prove that
\begin{equation}
\label{bound_cost}
{\cal C}[P(a)]\le {\cal C}.
\end{equation}
Given the distribution $P_\alpha(a)\equiv \alpha P(a)+(1-\alpha) P_M(a)$
with $\alpha\in[0,1]$, we have that
\begin{equation}
\label{ineq_deriv}
\left. \frac{d H_{P_\alpha(a)}(K|Y) }{d\alpha}\right|_{\alpha=0}\le 0,
\end{equation}
as $P_M(a)$ maximizes the conditional entropy. This equation implies Eq.~(\ref{bound_cost}), 
which can be seen by explicitly performing the derivation of the conditional entropy. 
Thus, from Eqs.~(\ref{Lbar},\ref{bound_cost}), we have that $\bar L[P(a)]\le{\cal C}+1$. 
Since this inequality holds for every $P(a)$, we have that $L(a)\le{\cal C}+1$.

\begin{figure}
\epsfig{figure=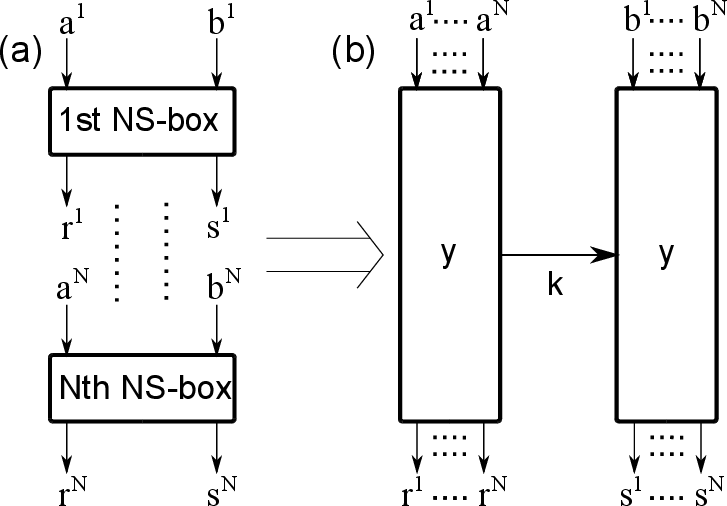,width=7.cm}
\caption{(a) $N$ identical nonsignaling boxes. On one side, Alices chooses the inputs $(a^1,\dots,a^N)$ 
and gets the outcomes $(r^1,\dots,r^N)$. On the other side, Bob chooses the inputs 
$(b^1,\dots,b^N)$ and gets the outcomes $(s^1,\dots,s^N)$. (b) Simulation of the $N$ 
nonsignaling boxes through a shared stochastic variable $y$ and the communication of the variable $k$.}
\label{fig2}
\end{figure}

If block-coding of many parallel instances of $k$ is employed, the minimal expected 
length $L(a)$ per instance turns out to be equal to $\cal C$ in the asymptotic
limit of infinite instances. However, block-coding of $k$ is not the most general
compression method in the case of
a parallel simulation of NS-boxes. A more general protocol simulating $N$ NS-boxes, 
which is schematically represented in Fig.~\ref{fig2}a, is as follows.
The $i$th box has input $a^i$ and 
outcome $r^i$ on one side (Alice side), and input $b^i$ and outcome $s^i$ on the 
other side (Bob side). Alice chooses the input $(a^1,\dots,a^N)\equiv \vec a$ and 
gets the outcome $(r^1,\dots,r^N)\equiv \vec r$. Similarly, Bob chooses the input 
$(b^1,\dots,b^N)\equiv\vec b$ and gets the outcome $(s^1,\dots,s^N)\equiv\vec s$.
Hereafter, we always use a superscript as a label of the NS-boxes.
The parallel simulation of $N$ NS-boxes is the same as for a single box, with 
$a$, $b$, $r$ and $s$ replaced by $\vec a$, $\vec b$, $\vec r$ and $\vec s$. 
This more general scheme produces a global variable $k$ with a probability depending 
on the overall input $\vec a$. The protocol exactly simulates the $N$ boxes if 
\be
\label{par_cond_sim}
\sum_k\int dy P(\vec s|y\,\vec b\,k)P(k,\vec r|y\,\vec a)\rho(y)= 
\prod_i P(r^i,s^i|a^i,b^i).
\ee
Each parallelized protocol has $N$ as a free parameter.
The asymptotic communication cost of the protocol is defined as
$\lim_{N\rightarrow\infty}{\cal C}^{par}/N\equiv {\cal C}^{asym}$, where ${\cal C}^{par}$
is the communication cost of the parallelized simulation. In this case, the maximization
in Eq.~(\ref{comm_cost}) is performed over the space of joint input distributions 
$P(a_1\dots a_N)$. We define the {\it nonlocal capacity} of the NS-box
as the minimum of ${\cal C}^{asym}$ among the parallelized protocols. The 
nonlocal capacity is denoted by ${\cal C}_{nl}^{asym}$.

\section{Computation of nonlocal capacity as a convex optimization problem}

Our task is to reduce the computation of ${\cal C}_{nl}^{asym}$ to the minimization
of a functional over a suitable space of distributions. Let us define this
space. \newline
{\bf Definition 1.} Given a nonsignaling box with conditional probability $P(r,s|a,b)$,
the set ${\cal V}$ contains any conditional probability $\rho(r\,{\bf s}|a)$ 
over $r$ and the sequence ${\bf s}=(s_1,\dots,s_M)$ whose marginal distribution of 
$r$ and the $m$-th variable is the distribution $P(r,s|a,b=m)$.
In other words, the set ${\cal V}$ contains any $\rho(r\,{\bf s}|a)$ satisfying the 
constraints
\begin{equation}\label{constraints}
\sum_{{\bf s},s_m=s} \rho(r\,{\bf s}|a)=P(r,s|a,b=m),
\end{equation}
where the summation is over every index in $\bf s$ except the $m$-th one, which is set 
equal to $s$. The subscript ``${\bf s}, s_m=s$'' means that the summation is done
over $\vec s$ with the constraint that the component $s_m$ is taken equal to $s$,
that is,
\begin{equation}
\sum_{{\bf s},s_m=s}=\sum_{{\bf s}} \delta_{s_m,s},
\end{equation} 
$\delta_{a,b}$ being the Kronecker delta.
\newline
\vspace{2mm}
\newline
The set $\cal V$ is surely non-empty. Indeed, the distribution 
$\rho(r\,{\bf s}|a)=P(s_1|r\,a\,b=1)\times \dots\times P(s_M|r\,a\,b=M)P(r|a)$
is an element of ${\cal V}$.
Note that $\rho(r\,{\bf s}|a)$ can be defined only if the first nonsignaling 
condition~(\ref{ns-conds}) is satisfied. The conditional probability
$\rho(r\,{\bf s}|a)$ defines a new box with a single input, $a$. We call
this box `HV-box', where HV stands for `hidden variable'. Indeed, this box
gives simultaneously the outcomes for every query $b$ of Bob, whereas
this information is partially hidden in a query of the original 
NS-box. 

\begin{center}
\begin{figure}
\epsfig{figure=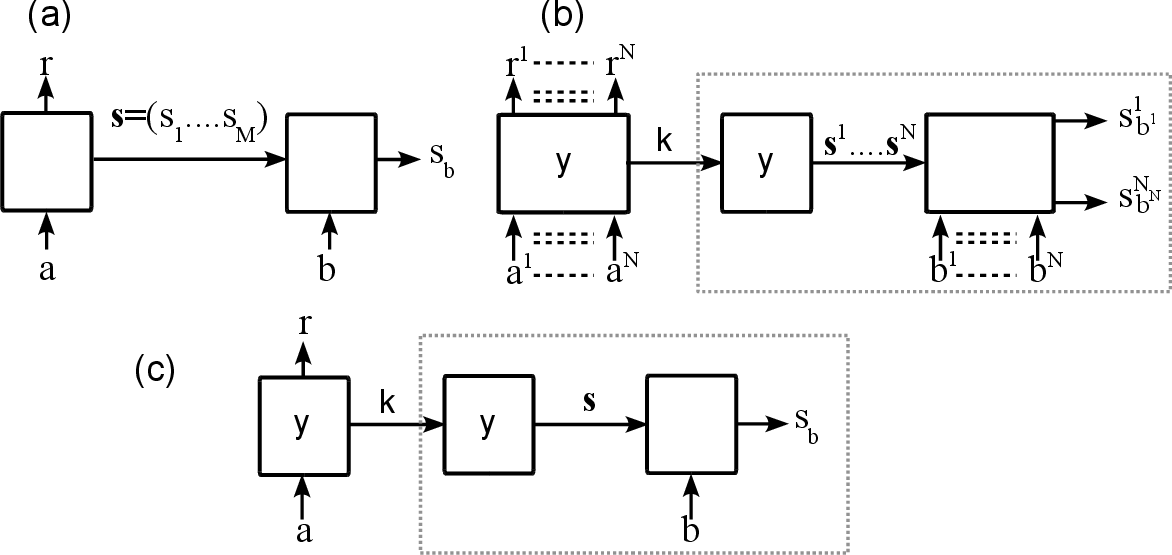,width=12.cm}
\caption{(a) Master protocol using a HV-box $\rho(r,{\bf s}|a)\in{\cal V}$ 
for simulating a NS-box. Alice sends $\bf s$, generated according to
$\rho(r,{\bf s}|a)\in{\cal V}$. Bob generates the outcome $s_b$.
(b) Child protocol using the simulation of $N$ HV-boxes by
employing shared randomness and communication. First, $N$
identical HV-boxes $\rho(r^1,{\bf s}^1|a^1),\dots,\rho(r^N,{\bf s}^N|a^N)\in{\cal V}$
are simulated as follows. Alice, who chooses $a^1,\dots,a^N$ and generates
$r^1,\dots,r^N$, sends $k$, enabling Bob to generate the variables 
${\bf s}^1,\dots,{\bf s}^N$ of $N$ instances according to the distributions
$\rho(r^i,{\bf s}^i|a^i)\in\cal V$. The two parties share the
random variable $y$.
Finally, Bob generates the outcomes $s^1_{b^1},\dots,s^N_{b^N}$,
as done in the master protocol for a single instance.
(c) One-shot child protocol. A single HV-box is simulated.}
\label{fig3}
\end{figure}
\end{center}

There is a trivial protocol that simulates a NS-box through its HV-box. 
Using the same terminology introduced in Ref.~\cite{montina1} in the context of channels,
we introduce the following protocol (Fig.~\ref{fig3}a) 
that simulates a NS-box through one of its HV-boxes.\newline
{\bf Master protocol.} 
Alice generates the outcome $r$ and the array $\bf s$ according to a conditional 
probability $\rho(r\,{\bf s}|a)\in{\cal V}$. Then, she sends $\bf s$ to Bob. 
Finally, Bob chooses the input $b$ and gives the outcome $s=s_b$.
\vspace{2mm}
\newline
It is obvious that $r$ and $s$ are generated according to the conditional
probability $P(r,s|a,b)$.

\onecolumngrid
\twocolumngrid

Through the procedure discussed in Ref.~\cite{montina} and used in 
Ref.~\cite{montina1} for quantum channels, we show that it is possible to 
turn the master protocol into a child protocol (Fig.~\ref{fig3}b) for
parallel simulations whose asymptotic communication cost is the capacity
of the channel $a\rightarrow {\bf s}$ associated to the conditional probability 
$\rho({\bf s}|a)\equiv\sum_r \rho(r\,{\bf s}|a)$. Let us recall that a channel
$x_1\rightarrow x_2$ is identified by a conditional probability distribution
$\rho(x_2|x_1)$ and its capacity is the maximum of the mutual information between 
$x_1$ and $x_2$ over the space of probability distributions $\rho(x_1)$~\cite{cover}. 
Let us denote by $C(x_1\rightarrow x_2)$ the capacity of the channel 
$x_1\rightarrow x_2$. The procedure in Ref.~\cite{montina}
is based on the Reverse Shannon theorem~\cite{rev_s}. 
Using the single-shot version of the reverse Shannon theorem~\cite{harsha}, we
also show that there is a single-shot simulation of a NS-box (Fig.~\ref{fig3}c), 
with associated HV-box $\rho(r,{\bf s}|a)$, whose communication cost is equal to 
$C(a\rightarrow {\bf s})$ plus an additional term scaling as
$\log C(a\rightarrow {\bf s})$.

Let us first prove the following.
\newline
{\bf Lemma 1.} 
Multiple instances of identical HV-boxes $\rho(r,{\bf s}|a)$ can be simulated 
in parallel through local randomness and communication with asymptotic communication 
cost ${\cal C}^{asym}$ equal to the capacity of the channel 
$\rho({\bf s}|a)\equiv\sum_r\rho(r\,{\bf s}|a)$. That is,
\begin{equation}
{\cal C}^{asym}=C(a\rightarrow{\bf s})
\end{equation}
The communication is established from Alice to Bob. Alice chooses an input $a^i$ 
and gets an outcome $r^i$ in each instance $i$, whereas Bob gets the outcomes 
${\bf s}^i$. The two parties can use shared random variables. Furthermore, there 
is a single-shot simulation of a HV-box with communication cost $\cal C$
such that
\begin{equation}
\label{bounds_single_shot}
{\cal C}^{asym}\le {\cal C}\le {\cal C}^{asym}
+2\log_2[{\cal C}^{asym}+1]+2\log_2e.
\end{equation}
\newline
{\it Proof.} The simulation is as follows. The $i$th instance has
input $a^i$ and outcome $r^i$ on Alice's side, and outcome ${\bf s}^i$ on Bob's
side. The outcomes are generated with conditional probability 
$\rho(r^i,{\bf s}^i|a^i)$. Alice chooses the inputs 
$a^1,\dots,a^N$. Then she sends Bob an amount of information
that allows Bob to generate the variables ${\bf s}^1,\dots,{\bf s}^N$ according
to the conditional probability $\rho({\bf s}^i|a^i)$. The reverse Shannon theorem~\cite{rev_s}
states that there is a protocol for this task with asymptotic communication cost 
equal to the capacity of the channel $\rho({\bf s}|a)$, provided
that Alice and Bob share some stochastic variable, say $\chi$. It is always possible
to have a deterministic protocol, so that the outcomes ${\bf s}^1,\dots,{\bf s}^N$ 
are uniquely determined by $\chi$ and the communicated information. Since $\chi$ is
shared with Alice, Alice knows Bob's outcomes. Thus, Alice generates
her outcomes $r^1,\dots,r^N$ according to the conditional probability 
$\rho(r^i|a^i\,{\bf s}^i)\equiv \rho(r^i\,{\bf s}^i|a^i)/\rho({\bf s}^i|a^i)$.
The overall set of outcomes is generated  according to the joint distribution
$\rho(r^i,{\bf s}^i|a^i)$. The last statement of the lemma has a similar proof and uses
the result in Ref.~\cite{harsha}.
The single-shot version of the reverse Shannon theorem proved in Ref.~\cite{harsha} 
states
that a single-shot simulation of the channel $a\rightarrow {\bf s}$ can be performed
with a communication cost $\cal C$ satisfying the inequalities~(\ref{bounds_single_shot}).
$\square$
\newline
Lemma~1 directly implies the following.
\newline
{\bf Lemma 2.}
Identical NS-boxes can be simulated in parallel with an asymptotic communication cost 
${\cal C}^{asym}$
equal to the capacity of the channel $\rho({\bf s}|a)\equiv\sum_r\rho(r\,{\bf s}|a)$,
where $\rho(r\,{\bf s}|a)\in{\cal V}$ is an associated HV-box. The parallel protocol
is obtained by using a parallel simulation of HV-boxes that employs shared randomness and
communication. The overall protocol, called {\it child protocol}, is represented
in Fig.~\ref{fig3}b. Furthermore, there is a single-shot simulation of a NS-box with
communication cost $\cal C$ satisfying the inequalities~(\ref{bounds_single_shot})
(Fig.~\ref{fig3}c).
\vspace{2mm}
\newline
{\it Proof.} 
This is a trivial consequence of Lemma~1. Indeed, a NS-box $P(r,s|a,b)$ can 
be simulated by a master protocol through an associated HV-box $\rho(r,{\bf s}|a)$,
which can be simulated with asymptotic communication cost equal to the 
capacity of the channel $\rho({\bf s}|a)$ and with single-shot communication cost
$\cal C$ satisfying constraints~(\ref{bounds_single_shot}).
$\square$

Let us define the quantity
\begin{equation}
\label{def_D}
{\cal D}\equiv\min_{\rho(r,{\bf s}|a)\in{\cal V}} \max_{P(a)} I({\bf S};A)
\equiv \min_{\rho(r,{\bf s}|a)\in{\cal V}} C(a\rightarrow {\bf s})
\end{equation}
associated to a NS-box $P(r,s|a,b)$,
where $I({\bf S};A)$ is the mutual information between the stochastic variables
$\bf s$ and $a$ and $C(a\rightarrow {\bf s})$ the capacity of the channel
$\rho({\bf s}|a)\equiv \sum_r\rho(r,{\bf s}|a)$.
Clearly, $\cal D$ is the minimum of the capacity of the channels
$a\rightarrow {\bf s}$ that are the marginals of the channels
$a\rightarrow r,{\bf s}$ in the set $\cal V$. Lemma~2 implies that the
nonlocal capacity of the NS-box $P(r,s|a,b)$ satisfies the inequality
\begin{equation}
{\cal C}_{nl}^{asym}\le {\cal D}.
\end{equation}
The next main task is to prove that the optimal parallel protocol simulating identical 
NS-boxes is given by a child protocol (schematized in Fig.~\ref{fig3}b) employing a 
simulation of parallel HV-boxes. The proof is a consequence of the 
data-processing inequality~\cite{cover}, which implies that 
${\cal C}_{nl}^{asym}\ge \cal D$, and therefore
\begin{equation}
\label{main_identity}
{\cal C}_{nl}^{asym}={\cal D}.
\end{equation}

Let us first consider a similar proof for the single-shot case, which is
less intricate and elucidates the key ideas used in the proof of the main theorem.
Namely, we show that ${\cal C}_{nl}\ge {\cal D}$. 
\newline
{\bf Theorem 1.}
The communication complexity ${\cal C}_{nl}$ of a NS-box $P(r,s|a,b)$ satisfies
the bounds
\begin{equation}
\label{ineqs_single_shot}
{\cal D}\le {\cal C}_{nl}\le{\cal D}+2\log_2({\cal D}+1)+2\log_2e.
\end{equation}

In few words, the procedure used 
in the proof of the Theorem~1 is as follows. Given an optimal 
protocol with communication cost ${\cal C}={\cal C}_{nl}$, we build a HV-box such 
that the associated capacity $C(a\rightarrow {\bf s})$ is not greater than ${\cal C}$.
This and the definition of ${\cal D}$ imply that ${\cal C}_{nl}\ge {\cal D}$.
\newline
{\it Proof.}
The second inequality is consequence of Lemma~2. Let us prove the first
inequality. Let $P(s|y\, b\, k)$, $P(k,r|y\, a)$ and $\rho(y)$ be the probability
distributions defining the optimal single-shot protocol. Thus, the associated 
communication cost $\cal C$ is equal to the communication complexity ${\cal C}_{nl}$.
Now, let us show that there is a channel $\rho(r\,{\bf s}|a)\in{\cal V}$
such that the capacity of the reduced channel $\rho({\bf s}|a)$ is not greater
than ${\cal C}_{nl}$. The channel is
$$
\rho(r\,{\bf s}|a)\equiv \int dy\sum_k \left[\prod_b P(s_b|y\, b\, k)\right] P(k,r|y\, a)\rho(y)
$$
By construction, the distribution $\rho(r\,{\bf s}|a)$ is in the set $\cal V$. 
Indeed, 
\begin{equation}
\sum_{{\bf s},s_b=s} \rho(r\,{\bf s}|a)=
\int dy \sum_{k,y} P(s|y\, b\, k) P(k,r|y\, a)\rho(y),
\end{equation}
which is reduced to Eq.~(\ref{constraints}) in view of Eq.~(\ref{cond_sim}).
The stochastic variables $a$, $\bf s$ and $k$ satisfy the Markov chain
$a\xrightarrow{y} k \xrightarrow{y} {\bf s}$, where the label
above that arrows denotes the shared variable $y$. From the data-processing
inequality~\cite{cover}, we have that
$$
\max_{P(a)}I({\bf S};A)\le \max_{P(a)}I(K;A|Y),
$$
where $I(K;A|Y)$ denotes the conditional mutual information between $k$ and $a$
given $y$. The mutual information between two variables is always less than
or equal to the entropy of each variable. Thus, 
$\max_{P(a)}I(K;A|Y)\le \max_{P(a)} H_{P(a)}(K|Y)$ and, from the definition
of communication cost, we have that
\begin{equation}
\label{data-proc_ineq}
\max_{P(a)}I({\bf S};A) \le \max_{P(a)}I(K;A|Y)\le
\max_{P(a)} H_{P(a)}(K|Y)={\cal C}={\cal C}_{nl}.
\end{equation}
Finally, from the definition of $\cal D$ and the fact that $\rho(r\,{\bf s}|a)\in{\cal V}$, 
we have that
$$
{\cal D}\le {\cal C}={\cal C}_{nl}.
$$
$\square$

The inequalities~(\ref{ineqs_single_shot}) provide tight lower and upper bounds
on ${\cal C}_{nl}$ and establish that the single-shot communication complexity
is equal to $\cal D$ up to an additional term scaling as the logarithm
of $\cal D$. As stated in the next theorem, the additional cost disappears
in the case of parallel simulations and the strict Eq.~(\ref{main_identity})
holds. The proof is more intricate and needs some further final efforts. 
As we said, a parallel protocol simulating $N$ NS-boxes is described by the conditional 
probabilities $P(\vec s|y\,\vec b\,k)$, $P(k,\vec r|y\,\vec a)$ and $\rho(y)$. The 
components of the sequences $\vec r=(r^1,\dots,r^N)$, $\vec s=(s^1,\dots,s^N)$, 
$\vec a=(a^1,\dots,a^N)$ and $\vec b=(b^1,\dots,b^N)$ are the inputs and outcomes
of each NS-box. Adapting the construction used in the proof of theorem~1 to the
parallel case, we build a multivariate HV-box 
$\rho(\vec r\,\vec {\bf s}|\vec a)=\rho(r^1 \dots r^N\,{\bf s}^1\dots{\bf s}^N|a^1\dots a^N)$ 
associated to the overall collection of NS-boxes, where ${\bf s}^i=(s_1^i,\dots,s_M^i)$.
The multivariate distribution is built so that the marginal distribution of $r^i$
and $s^i_b$ given $a^i$ is equal to $P(r^i,s^i_b|a^i,b)$, that is,
\begin{equation}
\label{constr_par}
\sum_{\vec {\bf s},\vec r,r^i=r,s^i_b=s} 
\rho(\vec r\,\vec {\bf s}|\vec a) \equiv \rho(r^i=r,s_b^i=s|\vec a)=
P(r,s|a^i,b),
\end{equation}
where  $\rho(r^i,s_b^i|\vec a)$ is the conditional marginal distribution of the
variables $r^i$ and $s_b^i$ given $\vec a$. Let us denote by ${\cal V}^{par}$
the set of multivariate HV-boxes satisfying this property on the marginals. The main key 
ingredient used in the proof of the next theorem is the equality
\begin{equation}
\label{key2}
N \times \min_{\rho(r,{\bf s}|a)\in{\cal V}} C(a\rightarrow {\bf s})=
\min_{\rho(\vec r,\vec {\bf s}|\vec a)\in{\cal V}^{par}} C(\vec a\rightarrow \vec{\bf s}).
\end{equation}
In particular, if $\rho(r,{\bf s}|a)$ minimizes $C(a\rightarrow {\bf s})$ in the set 
$\cal V$, then the factorized distribution 
\begin{equation}
\rho(\vec r,\vec {\bf s}|\vec a)=\prod_i \rho(r^i,{\bf s}^i|a^i),
\end{equation}
minimizes $C(\vec a\rightarrow\vec{\bf s})$ in the set ${\cal V}^{par}$.
It is clear that 
\begin{equation}
N \times \min_{\rho(r,{\bf s}|a)\in{\cal V}} C(a\rightarrow {\bf s})\ge
\min_{\rho(\vec r,\vec {\bf s}|\vec a)\in{\cal V}^{par}} C(\vec a\rightarrow \vec{\bf s}),
\end{equation}
as the capacity of a factorized channel $\rho(\vec {\bf s}|\vec a)=
\prod_{i=1}^N\rho(r^i,{\bf s}^i|a^i)\in{\cal V}^{par}$ is equal to the sum
of the capacity of the channels $\rho(r^i,{\bf s}^i|a^i)\in{\cal V}$.
To prove Eq.~(\ref{key2}), it is sufficient to show that the
minimum at the right-hand side is attained by a factorized function.
The proof is quite technical and is provided in Appendix~\ref{app3}. 
\newline
{\bf Theorem 2.} The nonlocal capacity ${\cal C}_{nl}^{asym}$ of an NS-box $P(r,s|a,b)$ 
is the minimum of the capacity of the channel $\rho({\bf s}|a)$ over the space $\cal V$
of associated HV-boxes. That is,
\be\label{nonlocal_cap}
{\cal C}_{nl}^{asym}={\cal D}.
\ee
\newline
{\it Proof.}
The inequality ${\cal C}_{nl}^{asym}\le {\cal D}$ is a consequence of Lemma~2. 
Let us prove that ${\cal C}_{nl}^{asym}\ge {\cal D}$. Let ${\cal C}^{par}$ be 
the communication cost of the optimal parallel protocol simulating $N$ NS-boxes. 
Thus, from the definition of nonlocal capacity, we have that
$$
\lim_{N\rightarrow\infty} {\cal C}^{par}/N={\cal C}_{nl}^{asym}.
$$
The protocol is defined by the conditional probabilities $P(k,\vec r|y\,\vec a)$ 
and $P(\vec s|y\,\vec b\,k)$ satisfying constraint~(\ref{par_cond_sim}).
Through a procedure described in Appendix~\ref{app1}, it 
is possible to build a multivariate HV-box with conditional probability 
\be\label{par_master_prot}
\rho(r^1\dots r^N\,{\bf s}^1\dots{\bf s}^N|a^1\dots a^N)=
\rho(\vec r\,\vec {\bf s}|\vec a)
\ee
over $r^i$ and the sequences ${\bf s}^i=(s_1^i\dots s_M^i)$
so that the following properties are satisfied.
\begin{enumerate}
\item The capacity of the channel $\rho(\vec {\bf s}|\vec a)$ is
smaller than or equal to ${\cal C}^{par}$. That is,
\begin{equation}
\label{data-proc_ineq2}
C(\vec a\rightarrow \vec {\bf s}) \le  {\cal C}^{par}.
\end{equation}
\item The marginal distribution of the variables $r^i$ and $s_b^i$ is equal to
$P(r^i,s_b^i|a^i,b)$, that is, constraints~(\ref{constr_par}) are satisfied.
\end{enumerate}
Ineq.~(\ref{data-proc_ineq2}) is similar to Ineq.~(\ref{data-proc_ineq}). The 
proof is identical and uses the data-processing inequality~\cite{cover} 
(see Appendix~\ref{app1}). From Eq.~(\ref{key2}), we have that 
\begin{equation}
N \times \min_{\rho(r,{\bf s}|a)\in{\cal V}} C(a\rightarrow {\bf s})
\le C(\vec a\rightarrow \vec {\bf s}) \le  {\cal C}^{par}.
\end{equation}
In the limit $N\rightarrow\infty$, we obtain
\begin{equation}
{\cal D}=
\min_{\rho(r,{\bf s}|a)\in{\cal V}} C(a\rightarrow {\bf s}) \le
\lim_{N\rightarrow\infty}\frac{{\cal C}^{par}}{N}={\cal C}^{asym}_{nl}.
\end{equation}
$\square$
\newline
This theorem reduces the evaluation of the nonlocal capacity of a NS-box to the computation 
of the quantity $\cal D$, defined by Eq.~(\ref{def_D}) as the minimum of the capacity
$C(a\rightarrow{\bf s})$ over the convex set $\cal V$. As the mutual information $I({\bf S};A)$
is convex in $\rho({\bf s}|a)$ and the maximum over a set of convex functions is
a convex function~\cite{boyd}, the capacity $C(a\rightarrow{\bf s})$ is convex
in $\rho({\bf s}|a)$. Thus, the computation of $\cal D$ is a convex optimization
problem, which is the main advantage of the presented method. Indeed, convexity
implies that every local minimum is a global minimum.
A different formulation of the problem has been introduced in Ref.~\cite{aref}, but it 
does not have the form of a convex minimization problem. This makes it harder to find 
the global minimum.

It is worth to note that the capacity $C(a\rightarrow{\bf s})$ does not have
generally an explicit analytic form and is not necessarily differentiable 
everywhere. This makes it harder to use standard methods of convex optimization~\cite{boyd}.
However, provided that the optimal distribution $P(a)$ maximizing the mutual 
information $I({\bf S};A)$ is known, the dual form of our optimization problem
is a geometric programming problem. This has been shown for the related problem
of computing the communication complexity of quantum channels~\cite{montina4,monti-wolf}.
Geometric programming is an extensively studied class of nonlinear optimization 
problems and can be solved by robust and very efficient algorithms~\cite{boyd2,chiang}.
A commercial implementation is provided by the MOSEK package (see http://www.mosek.com).
Even if $P(a)$ is not known, the minimization 
\begin{equation}
\label{just_min}
\min_{\rho(r,{\bf s}|a)} I({\bf S};A)\equiv C_-
\end{equation}
with any fixed $P(a)$ provides a lower bound on the nonlocal capacity, as a consequence 
of the minimax theorem~\cite{montina4,monti-wolf}. Furthermore, if $P(a)\ne 0$ for
every input $a$, then $\min_{\rho(r,{\bf s}|a)}I({\bf S};A)$ is different
from zero only and only if the correlations are nonlocal. Indeed, if the correlations
are local, $C_-$ is obviously equal to zero for every $P(a)$. Conversely, if 
there is a $\rho(r,{\bf s}|a)$ and a $P(a)\ne 0$ for every $a$ such that $I({\bf S};A)=0$,
then $I({\bf S};A)=0$ for every $P(a)$ and ${\cal D}$ is equal to zero. Thus, if we are interested
to discriminate between local and nonlocal correlations, we can fix $P(a)$,
for example by taking a uniform distribution, and solve the minimization~(\ref{just_min}).
A specialized numerical method that is particularly efficient for this problem and computes
also the optimal $P(a)$ will be discussed elsewhere. A similar method for computing
the communication complexity of quantum channels is discussed in Ref.~\cite{montina5}.

As the number of variables defining the probability distribution $\rho(r,{\bf s}|a)$
scales exponentially with respect to the number of Bob's measurements, our method
displays an exponential computational cost. Thus, it does not provide a better
scaling complexity than the computation of the distance from the nonlocal polytope.
However, numerical experiments show a speed difference of many decades between the 
two methods. Our method can solve a problem with 20 measurements in few minutes
or even few seconds, whereas the computation of the distance turns out to be very 
time-demanding even with 6 measurements. This difference is relevant if one
needs to test many different experimental configurations even if the number
of measurements is relatively small. Furthermore, the dual form of our optimization
problem displays very interesting properties, as stressed in 
Refs.~\cite{montina4,monti-wolf}. First, the number of dual variables scales
linearly with the size of the input of the problem, that is, with the number of 
variables defining the conditional probability $P(r,s|a,b)$. Second, although
the number of constraints grows exponentially, they are completely independent of 
$P(r,s|a,b)$. Thus, given a feasible point of the dual constraints, the computation
of a lower bound for \emph{every} nonlocal correlation has a linear computational
cost. This feature is employed in Ref.~\cite{sasha} to efficiently compute the setting 
of measurements maximizing the nonlocal capacity and, thus, providing the highest
degree of nonlocality.

Finally, it is worth to note that the distribution $\rho(r,{\bf s}|a)$ solving
the minimization problem is equal to zero for most of the values of $r$ and
${\bf s}$. Indeed, the numerical simulations and theoretical arguments show
that the support of the distribution grows linearly with the size of the
problem. Thus, most of the computational effort is to determine where
the distribution is equal to zero. It is an open question if the support
can be determined efficiently or even analytically in some relevant cases. 
This problem is related to the determination of feasible points of
the dual constraints.
In Ref.~\cite{montina4}, we showed an example involving infinite measurements,
for which we found analytically a nontrivial feasible point, from which
we determined nontrivial lower bounds for the communication complexity.

\section{Relationship with communication complexity of channels}

There is a relationship between the nonlocal capacity of a NS-box and the communication
complexity of a channel in a general probabilistic theory and, under some conditions, 
the computation of the former can be reduced to the computation of the latter, which
requires less sophisticated algorithms~\cite{montina5}. Furthermore, the relationship
allows us to directly infer results on the nonlocal capacity from known results
on the communication complexity of channels. For example, the analytical solution
found in Ref.~\cite{montina1} provides an analitical solution for maximally entangled
qubits and measurements associated to planar Bloch vectors. The communication complexity 
of a channel has been defined in Ref.~\cite{montina1}. The central scenario
studied there is the process of state preparation, communication through a channel, and
subsequent measurement. This process is described by a conditional probability 
$P(s|a;b)$, where $a$ and $b$ are inputs chosen by the sender (Alice) and the receiver
(Bob), respectively, and $s$ is an outcome obtained by Bob. In a general abstract
setting, we will just assume that $P(s|a;b)$ is any conditional probability depending
on two spatially separated inputs. We call this object C-box, where C stands for
channel. In Ref.~\cite{montina1}, a C-box is called game
$\bf G$. The asymptotic communication complexity of 
a C-box is the minimal asymptotic communication cost of a parallel simulation
of many copies of the C-box (See Ref.~\cite{montina1} for details). Let us denote 
this quantity by ${\cal C}_{ch}^{asym}$ (denoted by ${\cal C}_{min}^{asym}$
in Ref.~\cite{montina1}).
\newline
{\bf Corollary 1.} The nonlocal capacity ${\cal C}_{nl}^{asym}$ of an NS-box $P(r,s|a,b)$ 
satisfies the inequalities
\be
\begin{array}{c}
\label{claim}
{\cal C}_{ch}^{asym}+\min_{a\,r}\log_2 P(r|a)\le {\cal C}_{nl}^{asym}\le {\cal C}_{ch}^{asym} \\
-\min_a \max_b I(R;S|a\,b),
\end{array}
\ee
where ${\cal C}_{ch}^{asym}$ is the asymptotic communication complexity of the C-box 
$P(s|r\,a;b)\equiv P(r,s|a,b)/P(r|a)$ with Alice's inputs $r$ and $a$,
and Bob's input $b$. 
\newline
The first inequality can be proved by using a procedure described in Sec.~IIIB of 
Ref.~\cite{montina3}, the second inequality follows from Theorem~2 and the theorem
proved in Ref.~\cite{montina1}. The proof of the Corollary is given in Appendix~\ref{app2}.
\newline
{\bf Corollary 2.} Let $P(r,s|a,b)$ be a NS-box implemented through a maximally 
entangled state of two pairs of $n$ qubits ($n$ ebits of entanglement).
The two parties perform projective measurements and they share the same
set of allowed measurements. 
Then,
\be
{\cal C}_{nl}^{asym}={\cal C}_{ch}^{asym}-n,
\ee
where ${\cal C}_{ch}^{asym}$ is the capacity of the associated C-box $P(s|r a;b)$. The 
C-box can be implemented through a quantum channel with capacity $n$ qubits; the receiver
can perform any measurement allowed in the NS-box case, whereas the sender
can prepare any state corresponding to the eigestates of the allowed measurements.
\newline
{\it Proof.} The corollary follows directly from Corollary~1. Indeed, the lower
and upper bounds in Ineqs.~(\ref{claim}) collapse to the same value, as 
$P(r|a)=2^{-n}$ and $I(R;S|a,b=a)=n$.

\section{Numerical example}

To illustrate the presented method with an example, let us consider the
case of two-qubits in the Werner state $\rho_\gamma=\gamma/2(|00\rangle+|11\rangle)
(\langle 00|+\langle 11|)+(1-\gamma)\mathbf{1}/4$. What is the critical 
value of $\gamma$, denoted by $\gamma_0$, below which the Werner state
becomes local? This question is particularly interesting from an experimental
point of view, as $\gamma_0$ provides the amount of noise that
makes a maximally entangled state local. The Werner state admits a local 
model for $\gamma<0.659$~\cite{acin} and it is nonlocal for $1/\sqrt2\le\gamma\le1$, 
as the CHSH inequalities are violated. In Ref.~\cite{vertesi}, V\'ertesi derived a 
family of Bell inequalities that are violated for $\gamma>0.7056$, which is slightly
below the bound $1/\sqrt2$. This family requires $465$ measurement settings
on each side. Thus, $\gamma_0$ is between about $0.659$ and $0.7056$.
Is it possible to derive a better upper bound on $\gamma_0$ with a much smaller
set of measurements? To answer this question, we have computed the
nonlocal capacity for a number of measurements up to $20$ by trying
a high number of different settings, such as highly symmetric settings 
and random configurations. Each computation of the nonlocal capacity requires 
just few seconds on a standard laptop for the considered maximal set of
measurements. We always found local correlations
for $\gamma\le1/\sqrt{2}$. For example, given a set of $13$ measurements
corresponding to Bloch vectors pointing to the faces, edges and vertices of 
a cube (2 opposite vectors for each measurement), we find that ${\cal C}_{nl}^{asym}$
is equal to zero for $\gamma\le1/\sqrt2$ and is well-approximated by 
the analytic expression $9/4(\gamma-1/\sqrt2)^2$ 
for $\gamma\in[1/\sqrt2,1]$, with a maximum error lower than $3\%$ for
$\gamma=1$.
Our calculations suggest that, for a reasonable number of 
measurements, the CHSH inequalities are the best solution for testing 
nonlocality of a singlet in presence of noise. In a recent paper~\cite{montina4},
we derived the dual optimization problem for the computation of
the communication complexity of quantum channels and we used it
to derive an analytical lower bound on the communication complexity
in the case of infinite measurements. A similar dual problem can
be used to derive a lower bound on the nonlocal capacity of Werner
states and, thus, an upper bound on $\gamma_0$. We are currently
studying the possibility of analytically deriving the exact value
of the nonlocal capacity in the case of infinite measurements by
using the dual formulation.

\section{Conclusions}

In conclusion, we have presented a method for evaluating the nonlocal
capacity of correlations, and provided a tight lower and upper bound on
the communication complexity in the 
single-shot case. The introduced measure of nonlocality can be used as an
alternative to Bell's inequalities for testing if some given theoretical or
experimental data display nonlocal correlations. Furthermore, this measure
provides an upper bound to the power of nonlocal correlations,
as an information-theoretic resource, in terms of classical communication. 
In a subsequent work, we will present an efficient numerical method for
evaluating the nonlocal capacity and will derive a dual optimization problem
which can help to solve the open problem concerning the range of $\gamma$ where
a Werner state is nonlocal. A similar dual problem was recently derived in 
Ref.~\cite{montina4} for the optimization problem introduced in
Ref.~\cite{montina1}.

\section*{Acknowledgments.} 
This work is supported by the Swiss National Science Foundation, 
the NCCR QSIT, the COST action on Fundamental Problems in Quantum Physics and
Hasler Foundation under the project number 14030 
"Information-Theoretic Analysis of Experimental 
Qudit Correlations". The authors wish to thank Arne Hansen for useful
comments and the careful reading of the manuscript.

\bibliography{biblio.bib}

\onecolumngrid

\appendix

\section{Minimizing the capacity with constraints on the marginals}
\label{app3}
Let us consider the set ${\cal V}^{par}$ of channels 
$a_1,\dots,a_N\rightarrow (x_1,y_1,\dots,x_N,y_N)$ with conditional
probability $\rho(\vec x,\vec y|\vec a)$ satisfying the constraints
\begin{equation}
\label{constr_margin}
\sum_{\vec x,\vec y} B_{k,i}(x_i,y_i)\rho(\vec x,\vec y|\vec a)=A_{k,i}(a_i)\;\;
\forall k=1,\dots,K,\;\; i=1,\dots,N,
\end{equation}
where $B_{k,i}$ and $A_{k,i}$ are real numbers. 
The constraints~(\ref{constr_par}) take this form, once we identify
$\vec{\bf r}$, $\vec {\bf s}$ and $k$ with $\vec x$, $\vec y$ and
$(r,s,b)$, respectively. 

Note that the constraints can be rewritten in the form
\begin{equation}
\label{constr_margin2}
\sum_{x_i,y_i} B_{k,i}(x_i,y_i)\rho(x_i,y_i|\vec a)=A_{k,i}(a_i)\;\;
\forall k=1,\dots,K,\;\; i=1,\dots,N,
\end{equation}
where $\rho(x_i,y_i|\vec a)$ are the marginal conditional distributions
of the variables $x_i$ and $y_i$. Furthermore, whereas $\rho(x_i,y_i|\vec a)$ can
depend on every element of the sequence $a_i,\dots,a_N$, the sum at the
left-hand side of Eq.~(\ref{constr_margin2}) depends only on $a_i$,
as the right-hand side depends only on $a_i$.
\newline
{\bf Theorem.} There is a factorized distribution 
$\rho_f(\vec x,\vec y|\vec a)=\prod_i \rho(x_i,y_i|a_i)$ such
that the capacity of the reduced channel $\vec a\rightarrow \vec y$
is minimal in ${\cal V}^{par}$.
\newline
{\it Proof.}
Let $\rho_m(\vec x,\vec y|\vec a)$ be a channel that minimizes the
capacity of the reduced channel $\rho_m(\vec y|\vec a)$ under
the constraints~(\ref{constr_margin2}). We denote the channel
$\rho_m(\vec y|\vec a)$ by $M$. Now, we build another
distribution that is factorized and minimal in the set ${\cal V}^{par}$
of constrained channels.
Let us take a factorized distribution
$\rho(\vec a)=\prod_i\rho(a_i)$ for the variable $\vec a$ and introduce
the probability distribution $\rho(\vec x,\vec y,\vec a)$. Note the
abuse of notation. We should introduce some index, such as $\rho_i(a_i)$,
for distinguishing different distributions. For the sake of simplicity,
we will distinguish two distributions from their argument, so that $\rho(a_1)$
and $\rho(a_2)$ are not meant to be the same function.
Let $\rho(x_i,y_i,a_i)$ be the marginal distributions of the
variables $x_i$, $y_i$ and $a_i$. The conditional probability
distributions $\rho(x_i,y_i|a_i)$ define the channels 
$a_i\rightarrow x_i,y_i$. By construction, the multivariate
channel
\begin{equation}
\rho_f(\vec x,\vec y|\vec a)\equiv\prod_i \rho(x_i,y_i|a_i)
\end{equation}
satisfies the constraints~(\ref{constr_margin2}), as the right-hand
side of the constraints only depend on one component $a_i$.
Let the reduced factorized channel
\begin{equation}
\label{factorized_form}
\rho_f(\vec y|\vec a)\equiv\prod_i \rho(y_i|a_i)
\end{equation}
be denoted by $F$.
 
Let us choose the distributions $\rho(a_1),\dots,\rho(a_N)$ so that the mutual 
information $I(y_i;a_i)$ is maximal in $\rho(a_i)$ for every $i$. 
Note that each conditional probability $\rho(x_i|a_i)$ depends on the
full set of distributions $\rho(a_1),\dots,\rho(a_N)$, apart from
$\rho(a_i)$. Thus, the maximizations of the functions $I(y_i;a_i)$ 
are not independent optimization problems. Anyway, the overall problem has
a solution and this choice of
$\rho(a_1),\dots,\rho(a_N)$ is always possible.

Thus, by definition of channel capacity~\cite{cover},
$I(y_i;a_i)$ is the capacity of the reduced channel $a_i\rightarrow y_i$,
say $C(a_i\rightarrow y_i)$.
Furthermore, the capacity of the channel $F$, say $C(F)$, is
\begin{equation}
C(F)=\sum_i C(a_i\rightarrow y_i).
\end{equation}

We denote the mutual information between the stochastic variables $\vec a$
and $\vec y$ with conditional probability $\rho_m(\vec y|\vec a)$ and
marginal distribution $\rho(\vec a)=\prod_{i=1}^N \rho(a_i)$ by $I_m(\vec y;\vec a)$.
Using the chain rule for the mutual information~\cite{cover}
\begin{equation}
I(x,y;z)=I(x;z)+I(y;z|x)
\end{equation}
and the fact that $I(a_i;a_{i'})=0$ for $i\ne i'$,
let us prove that
\begin{equation}
\label{ineq1}
I_m(\vec y;\vec a) \ge \sum_i I(y_i;a_i)=\sum_i C(a_i\rightarrow y_i)=C(F).
\end{equation}
Intuitively, this inequality says that the variable $\vec y$ contains less 
information about $\vec a$ if the conditional probability 
$\rho_m(\vec y|\vec a)$ is replaced by the factorized form~(\ref{factorized_form}),
provided that the marginal distribution $\rho(\vec a)$ is factorized.
Indeed, the factorized form $\rho_f(\vec y|\vec a)$  looses the correlations 
among the components of $\vec y$, and these correlations can contain 
extra-information about $\vec a$. Conversely, if $\rho(\vec a)$ is
not factorized, the factorized form $\rho_f(\vec y|\vec a)$ can increase
the information about $\vec a$ for majority vote reasons.
Let us prove this inequality for $N=2$. The general case can be 
proved recursively.
\begin{equation}
\begin{array}{c}
I_m(y_1\,y_2;a_1\,a_2)=I_m(y_1\,y_2;a_1)+I_m(y_1\,y_2;a_2|a_1)=
I_m(y_1\,y_2;a_1)+I_m(y_1\,y_2,a_1;a_2)-I_m(a_1;a_2)= \\
I_m(y_1\,y_2;a_1)+I_m(y_1\,y_2,a_1;a_2)=
I_m(y_1;a_1)+I_m(y_2;a_1|y_1)+I_m(y_2;a_2)+I_m(y_1\,a_1;a_2|y_2)\ge \\
I(y_1;a_1)+I(y_2;a_2).
\end{array}
\end{equation}
As the capacity $C(M)$ of the channel $M$ is the maximum of
the mutual information $I(\vec y;\vec a)$ with
respect to the whole space of distributions $\rho(\vec a)$,
Ineq.~(\ref{ineq1}) implies the inequality
\begin{equation}
C(M) \ge C(F).
\end{equation}
Since $C(M)$ is minimal under the constraints~(\ref{constr_margin}) 
and $\rho_f(\vec x,\vec y|\vec a)$ 
satisfies the constraints, the factorized channel 
$\rho_f(\vec x,\vec y|\vec a)$ 
also minimizes the capacity of the reduced channel 
$\vec a\rightarrow \vec y$ in the set ${\cal V}^{par}$.
$\square$

\section{Proof of Properties $1$ and $2$ in Theorem 2}
\label{app1}

The protocol is defined by the conditional probabilities $P(k,\vec r|y,\vec a)$, 
$P(\vec s|y\,\vec b\,k)$ and $\rho(y)$ satisfying the constraints~(\ref{par_cond_sim}).

First, we note that Bob does not need to generate the outcome of the $i$-th NS-box
with a probability depending on the inputs of the other boxes. If this independence
property is not satisfied by $P(\vec s|y\,\vec b\,k)$, it is always possible to replace 
Bob's protocol with one satisfying the property, without affecting Alice's protocol 
and, thus, the communication cost. Thus, we can safely
assume that $P(\vec s|y\,\vec b\,k)$ is factorized as follows
\be
P(s^1\dots s^N|y,b^1\dots b^N,k)=\prod_i P^i(s^i|y,b^i,k).
\ee
Let us introduce the conditional probability
\be
P^i(s_1,s_2,\dots,s_M|y,k)\equiv
\prod_{b=1}^M P^i(s_b|y,b,k),
\ee
We will concisely denote $P^i(s_1,s_2,\dots,s_M|y,k)$ by $P^i({\bf s}|y,k)$.
We use the probabilities $P^i({\bf s}|y,k)$ to build the conditional probabilities
\be
P({\bf s}^1,\dots,{\bf s}^N|y,k)=\prod_i P^i({\bf s}^i|y,k).
\ee
Finally, from this distribution and $P(k,\vec r|y,\vec a)$, we build the conditional probability
\be
\label{eq123}
\rho(\vec r,{\bf s}^1,\dots,{\bf s}^N|\vec a)= 
\sum_k\int dy\rho(y)P({\bf s}^1,\dots,{\bf s}^N|y,k)P(k,\vec r|y,\vec a).
\ee
As seen in the proof of Theorem~1, we obtain from the data processing inequality and
Eq.~(\ref{eq123}) that the capacity $C(\vec a\rightarrow\vec{\bf s})$ of the channel 
$\vec a\rightarrow({\bf s}^1,\dots,{\bf s}^N)$ satisfies the inequality
\begin{equation}
C(\vec a\rightarrow\vec{\bf s})\le \max_{P(\vec a)} I(K;A|Y)\le
\max_{P(\vec a)}H_{P(a)}(K|Y)={\cal C}^{par},
\end{equation}
which is Property~1.

By construction and Eq.~(\ref{par_cond_sim}), we also have that
the constraints~(\ref{constr_par}) are satisfied, i.e. Property~2.

\section{Proof of Corollary~1.}
\label{app2}

The first inequality can be proved by using a procedure described in Sec.~IIIB of 
Ref.~\cite{montina3}, where we showed that a protocol for simulating a maximally 
entangled state of $n$ qubits can be used to simulate the communication of $n$ qubits with an 
additional cost of $n$ classical bits. More generally, the additional cost is not
more than $-\min_{r\,a}\log_2 P(r|a)$. Let us prove it. Any protocol simulating $N$ 
NS-boxes is deterministic or can be made deterministic by introducing some additional 
random variables shared by Alice and Bob. This means that the two parties have
a common list noise realizations, say $y_1,y_2,\dots$. In the NS-box simulation,
Alice chooses $a$, but she cannot choose her outcome $r$, which is determined
by $a$ and the noise. Conversely, in the C-box simulation, Alice chooses both
$a$ and $r$ such that the conditional probability $P(s|r\; a;b)$ of Bob's
outcome in the C-box and NS-box simulations are identical. The C-box can
be simulated as follows. Alice starts reading the noise list from the first
element and stops at the value $y_i$ that generates the outcome $r$ chosen
by her. Then, she uses the communication procedure of the NS-box protocol
and, furthermore, she sends the index $i$. Finally, Bob uses the noise 
realization $y_i$ in the NS-box protocol. The additional cost is
not bigger than $-\min_{a,r}\log_2 P(r|a)$.

Now, let us prove the second inequality.
Given the NS-box $P(r,s|a,b)$, let $\cal V$ and ${\cal V}'$ be the space of associated 
HV-boxes $\rho(r {\bf s}|a)$ and distributions 
$\rho({\bf s}|r a)\equiv\rho(r {\bf s}|a)/\rho(r|a)$, respectively.
The chain rule~\cite{cover}
\be
I({\bf S};A)=I({\bf S};R,A)-I({\bf S};R|A)
\ee
for the mutual information, the definition of a HV-box and the data-processing 
inequality imply that
\be
I({\bf S};A)\le I({\bf S};R,A)-\min_a\max_b I(R;S|a b).
\ee
As the marginal distribution of $r$ given $a$ is the same for every distribution
$\rho(r {\bf s}|a)$ in $\cal V$, the maximization in Eq.~(\ref{nonlocal_cap})
giving the nonlocal capacity can be performed over the space 
$\rho({\bf s}|r a)\in {\cal V}'$. Thus, we have that
\be\label{first_ineq}
{\cal C}_{nl}^{asym}\le \min_{\rho({\bf s}|r a)\in{\cal V}'}\max_{P(a)} I({\bf S};R,A)
-\min_a \max_b I(R;S|a\,b).
\ee
We also have that
\be\label{second_ineq}
\min_{\rho({\bf s}|r a)\in{\cal V}'}\max_{P(a)} I({\bf S};R,A)\le
\min_{\rho({\bf s}|r a)\in{\cal V}'}\max_{P(r,a)} I({\bf S};R,A).
\ee
In Ref.~\cite{montina1}, we showed that the right-hand side is equal to 
${\cal C}_{ch}^{asym}$. Thus, the inequalities~(\ref{first_ineq},
\ref{second_ineq}) imply the second inequality.

\end{document}